\documentclass[a4paper]{jpconf}
\usepackage[icelandic,english]{babel}
\usepackage[T1]{fontenc}

\usepackage{graphicx}
\graphicspath{{./}{Graphics/}}

\usepackage{booktabs}
\usepackage{array} 
\newcolumntype{L}[1]{>{\raggedright\let\newline\\\arraybackslash\hspace{0pt}}m{#1}}
\newcolumntype{C}[1]{>{\centering\let\newline\\\arraybackslash\hspace{0pt}}m{#1}}
\newcolumntype{R}[1]{>{\raggedleft\let\newline\\\arraybackslash\hspace{0pt}}m{#1}}

\usepackage{custom}
\usepackage{url}
\begin{document}
\title{Repeat a physics course more than once: A diagnosis of the most frequent misconceptions in newtonian mechanics of first and second-year engineering students}

\author{F. Escalante$^{1}$}

\address{$^{1}$Departamento de Física, Universidad Católica del Norte, Avenida Angamos 0610, Casilla 1280, Antofagasta, Chile.}

\begin{abstract}
In this work, we study a sample of 600 engineering students in three consecutive physics courses of different levels, according to their curriculum. We focus on the most frequent students’ misconceptions in newtonian mechanics. It was administered the FCI test at the beginning of the academic semester in each course. We use the taxonomy of misconceptions probed in the FCI  test, and through this, we establish correlations between the number of attempts in each course, to explore if there are misconceptions that repeat in the students that are on more than one attempt-in each course. The data reported in this paper were analyzed using the method of dominant incorrect answers. The results show that there is no relevant difference between a student on his first attempt and a student on his third or more attempt, in each course, in terms of the most frequent misconceptions.
\end{abstract}

\section{Introduction}
In the early years of higher education, the students face physics courses, and one of the transversal topics in their curriculum is newtonian physics. 
A widely used tool to measure the understanding of the students in newtonian physics is the Force Concept Inventory (FCI)~\cite{FCI}, which in their original version it's composed of 29 multiple choice question split into six items:  kinematics, first Newton law, second Newton law, third Newton law, superposition principle and kinds of force, where each one of the alternatives its related to one or more misconceptions. This test has been modified, being their alternative version of a 30 multiple choice question test.

\begin{table*}[t]
 \centering
  \begingroup
  \caption{Taxonomy of misconceptions probed by the inventory with their corresponding code.}
    \label{tax}
\begin{tabular}{lll}
 \toprule
Section & Code & Misconception  \\
\hline
	&	K1	&	 Position-velocity undiscriminated	\\
Kinematics	&	K2	&	 Velocity-acceleration undiscriminated	\\
	&	K3	&	 Nonvectorial velocity composition	\\
	\hline
	&	I1	&	 Impetus supplied by ‘hit’	\\
	&	I2	&	 Loss/recovery of original impetus	\\
Impetus	&	I3	&	 Impetus dissipation	\\
	&	I4	&	 Gradual/delayed impetus build-up	\\
	&	I5	&	 Circular impetus	\\
	\hline
	&	AF1	&	 Only active agents exert forces	\\
	&	AF2	&	 Motion implies active force	\\
Active Force	&	AF3	&	 No motion implies no force	\\
	&	AF4	&	 Velocity proportional to applied force	\\
	&	AF5	&	 Acceleration implies increasing force	\\
	&	AF6	&	 Force causes acceleration to terminal velocity	\\
	&	AF7	&	 Active force wears out	\\
	\hline
Action/Reaction pairs	&	AR1	&	 Greater mass implies greater force	\\
	&	AR2	&	 Most active agent produces greatest force	\\
	\hline
	&	CI1	&	 Largest force determines motion	\\
Concatenation of Influences	&	CI2	&	 Force compromise determines motion	\\
	&	CI3	&	 Last force to act determines motion	\\
	\hline
Centrifugal Force	&	CF	&	 Centrifugal force	\\
\hline
Obstacles exert no force	&	Ob	&	 Obstacles exert no force	\\
\hline
	&	R1	&	 Mass makes things stop	\\
Resistance	&	R2	&	 Motion when force overcomes resistance	\\
	&	R3	&	 Resistance opposes force/impetus	\\
	\hline
	&	G1	&	 Air pressure-assisted gravity	\\
	&	G2	&	 Gravity intrinsic to mass	\\
Gravity	&	G3	&	 Heavier objects fall faster	\\
	&	G4	&	 Gravity increases as objects fall	\\
	&	G5	&	 Gravity acts after impetus wears down	\\

\bottomrule
\end{tabular}
\endgroup
\end{table*}

The students' understanding and failure of this topic, in their first-year physics courses, have been studied over the years~\cite{poutot, bao1, vonkorff, blas, bani, hoellwarth, hammer1, hammer2}, and in addition, it has been studied the capability to solve problems~\cite{rakkapao, stoen}. However, no literature relates the attempts, when repeating physics courses, and most frequent students’ misconceptions in newtonian mechanics topics.

Unlike other universities around the world, Chilean university students that approve a course on their first attempt, cannot repeat it to improve their marks or reinforce their background. Diverse studies in other knowledge areas deepen in the effects of repeating courses~\cite{casas,dibss,tafreschi,armstrong}, nevertheless there exist controversy regarding the effectiveness of repeating courses voluntarily as a student could improve their background knowledge but at the same time increase the student retention.

In this study, we analyze a sample of first and second-year Chilean Engineering students, focusing on their most frequent misconceptions in the subject of Newtonian mechanics, considering the number of attempts in which they take a course and their results in the FCI test. This study took as a sample, students from three courses of different levels (first, second and fourth semester of higher education) that address the topic of newtonian mechanics.

\begin{table*}[t]
  \centering
  \begingroup
  \caption{It present the 29 questions of the FCI test and each alternative whit their corresponding misconception taxonomy code.}
    \label{misconceptions}
\begin{tabular}{l*{5}{c}r}
\toprule
 & A & B & C & D & E \\
\hline
1 & G3 &  &  &  &  \\
2 & AR1 &  & Ob & AR1 &  \\
3 &  & G3 &  & G3 &  \\
4 & I5 &  & CF, CI2 & CF, I2, I5 & CF \\
5 & I3 & G4, G5, I3 & I3 &  & G2 \\
6 & CI3 &  & I2 & I4 & I2 \\
7 &  & CI3 & K3 &  &  \\
8 &  & I4 & I3 & I4 &  \\
9 & G1, Ob & Ob, I1 & I1 &  & G2 \\
10 & I5 &  & CF & CF, CI2 & CF \\
11 &  & AF1 &  & AR1, AR2 &  \\
12 & Ob & AF1 & G1 &  & AF3 \\
13 &  & AR1 & AR2 & AF1 & Ob \\
14 &  & AR1 & AR2 & AF1 & Ob \\
15 & AF1 & AF1 &  &  &  \\
16 & CI2 &  & I3 & G5, I3 &  \\
17 & AF6 & AF5, G4 & CI2 & G2 & G1 \\
18 & CI1 &  &  & AF1 & CI1, G1 \\
19 & CI1 &  &  & CI2 &  \\
20 & K2 & K1 & K1 & K1 &  \\
21 &  & K2 & K2 &  &  \\
22 & AF1 & I1 & I1 &  & I1 \\
23 & R1 & R1 & CI2 &  & G5, I3 \\
24 & I2 & CI3 & CI2 & I4 &  \\
25 & AF4 &  & AF7 & AF6 & AF7 \\
26 & I2 &  & CI3 & I2 & I2 \\
27 &  &  & I3 &  & I3 \\
28 & AF4 & R2 &  & R2 & R3 \\
29 & AF2, R1 & I3, R1 &  & I1 & I4 \\
\bottomrule
\end{tabular}
\endgroup
\end{table*}

\section{Context}
Chilean university students who fail a course must repeat it until it has been approved, nevertheless, the number of attempts they can repeat a course depends on the internal regulation of each university. The students of the sample of this study, belonging to the Universidad Católica del Norte, Chile, can repeat a course three times, but, in case of failure in their third attempt, it could summit to consideration of the career council a request for one last time only. This means that, in special cases, a student could repeat a course four times.

Usually, remedial plans are made between semesters to be able to pass a course that was failed~\cite{mcginn,bahr}. These remedial plans have a brief duration and there are intensive as to demand and times as they have two to four weeks’ duration.

A student who passes a course on his first attempt, it's promoted to the next course in his curriculum and must face it with the acquired knowledge in the previous course. This point is relevant when a student must continue with related courses, such as physics courses. The students of the sample have three courses related to classical mechanics.

In the first semester, the students have an introductory Physics course (Phys. 1). This is the first Physics course that the engineering and physics students face, designed from a college-level bibliography~\cite{college}. This is the first approach to Classical Mechanics, where content such as Physics and measurement, vectors, one and two dimensional kinematics, Newton laws, work and energy, linear momentum, and gravitation are revised. Although motion content is reviewed in two dimensions, circular motion is out. This traditional course is the basis that students will have to face the following physics courses, especially those related to Classical Mechanics, along with their curriculum.

In their second semester, the students face a second physics course (Phys. 2). There, they explore the contents of Classical Mechanics~\cite{serway, sears, tipler} using a mathematical background such as a one-semester introductory calculus course. The difficulty of problems asked compared to the Phys. 1 course, increases. It is important to note that the topics of fluids, oscillations and mechanic waves, and thermodynamics are out.

The third and last course associated with Classical Mechanics,  Phys. 3, is located in the fourth or fifth semester, according to the career. This course pretends that students can apply learned concepts in previous courses, to problem-solving, using vector calculus tools. From used bibliography~\cite{beer, hibbeler}, mechanical vibration is out.

\begin{table*}
  \caption{Statistic per course of Number of students (N), Mean score (M) and standard deviation (SD), considering their attempts in each course.}
  \center
\begin{tabular}{llll}
 \toprule
\multicolumn{4}{c}{Phys. 1} \\
\cline{1-4}
	& First	&	Second	&	Third o more \\ \hline
N	&	19	&	150	&	16 \\
M	&	8.1	&	8.1	&	8.1 \\
SD	&	3.3	&	3.2	&	2.4 \\
 \bottomrule
\multicolumn{4}{c}{Phys. 2} \\
\cline{1-4}
	& First	&	Second	&	Third o more \\ \hline
N	& 198	&	30	&	12 \\
M	& 10.4	&	9.7	&	8.9 \\
SD	& 4.5	&	3.6	&	1.4  \\
\bottomrule
\multicolumn{4}{c}{Phys. 3} \\
\cline{1-4}
& First	&	Second	&	Third o more \\ \hline
N	&	113	&	35	&	27 \\
M	&	11.4	&	10.8	&	12.9 \\
SD	&	4.5	&	4.9	&	4.7 \\
 \bottomrule
\end{tabular}
\label{statistics}
\end{table*}

\section{Metodology}

The FCI test, in their Spanish version, was administered to the three courses mentioned before (Phys. 1, Phys. 2, and Phys. 3), as a pre-test, at the beginning of the second semester of 2019. A total of 600 students answered the full test, of which 185 are from Phys. 1, 240 from Phys. 2, and 175 from Phys. 3.  As it shows in table \ref{statistics}, 81\% of the students in Phys. 1 course is on their second attempt. This is because the students of the sample didn't pass it on their first attempt (or first semester), so they must attempt again. For the first attempt, there are 82.5\% in Phys. 2 and 64.4\% in Phys. 3.

Based on the above, our goal is to study the students' misconceptions of these three courses in the FCI test, independently. Using the misconception taxonomy showed in table \ref{tax}, and through this, establish correlations between the number of attempts in each course, to explore if there are misconceptions that repeat in the students that are on more than one attempt-in each course. For this purpose, we use the dominant incorrect answer method~\cite{blas} to identify which misconceptions are more frequent.

Table~\ref{tax} shows the taxonomy of misconceptions with their corresponding code. Every question of the FCI is a multiple-choice question, and some answers are related to one or more misconceptions, as shown in table~\ref{misconceptions}. The dominant incorrect answer method concentrates on studying the wrong answers by students and reporting the dominant ones for each question \cite{bani}. This method establishes that, in each option (e.g. A, B, C, D, or E), if the percentage of the ratio between the number of selected answers associated with a determined misconception and the number of the total incorrect answers, in each question, is equal or greater to 50\%, then this answer, related to the respective option, is a dominant incorrect answer.

\section{Results}
The overall performance of the students in each course it's shown in table~\ref{statistics}. The mean score in Phys. 1 course is the same, independent of the number of attempts. The mean score obtained by students in Phys. 2 course, although varies, there exists a decrease according to the attempts, which means, in a first approximation, that the more times a student attempts the course, their performance decreases in the FCI. On the other hand, we can see that the score dispersion is widely in their first attempt and narrower in their third or more attempts.

In the Phys. 3 course, the situation is similar to Phys. 1 and Phys. 2. Although the mean score fluctuates, these values don't spread too much between them and their standard deviations are very similar between them.
In general, we can see that the average score in each course is low and it's below the `` entry threshold '' to newtonian physics~\cite{Savinainen}.

In tables~\ref{desgloseI}, \ref{desgloseM} and \ref{desgloseD} it shows the questions who has a misconception which value is equal or greater than 50\%, classified by their attempts.

It is important to note that the shown percentages in tables~\ref{desgloseI}, \ref{desgloseM} and \ref{desgloseD} corresponds to students who answer incorrectly. For example, in table~\ref{desgloseI} question 5 shows that adding up the percentages of the alternatives A, B, and C, of the misconception I3, obtains 100\% of students who answer incorrectly and choose one of these alternatives. This number corresponds to 19 students who are in Phys. 1 course on their first attempt. A similar situation occurs for the students in their third or more attempts. In the case of students on their second attempt, it has a 99.3\% which corresponds to 141 in 142 students who answer incorrectly and selected the alternatives A, B, or C.

Tables~\ref{desgloseI} shows that there are repeating questions. As an example, question 4 is related to misconceptions I5 and CF, and on the other hand, question 17 is related to misconceptions AF5 and G4, which in both cases we can found twice.

Also, we can see from table~\ref{desgloseM} that questions 11 and 17 repeat twice each one. Question 11 is related to the misconceptions AR1 and AR2, and question 17 with misconceptions AF5 and G4.

Finally, from table~\ref{desgloseD} we can see again that questions 11 and 17 repeat twice, and there are associated with the misconceptions AR1, AR2, AF5, and G4.

Of the 30 misconceptions described by Hestenes et al.~\cite{FCI} (table \ref{tax}), and according to the dominant incorrect answer method,  17 misconceptions are present in Phys. 1 course and 15 in Phys. 2 and Phys. 3 courses.

\begin{figure}
    \centering
    \includegraphics[width=0.6\textwidth]{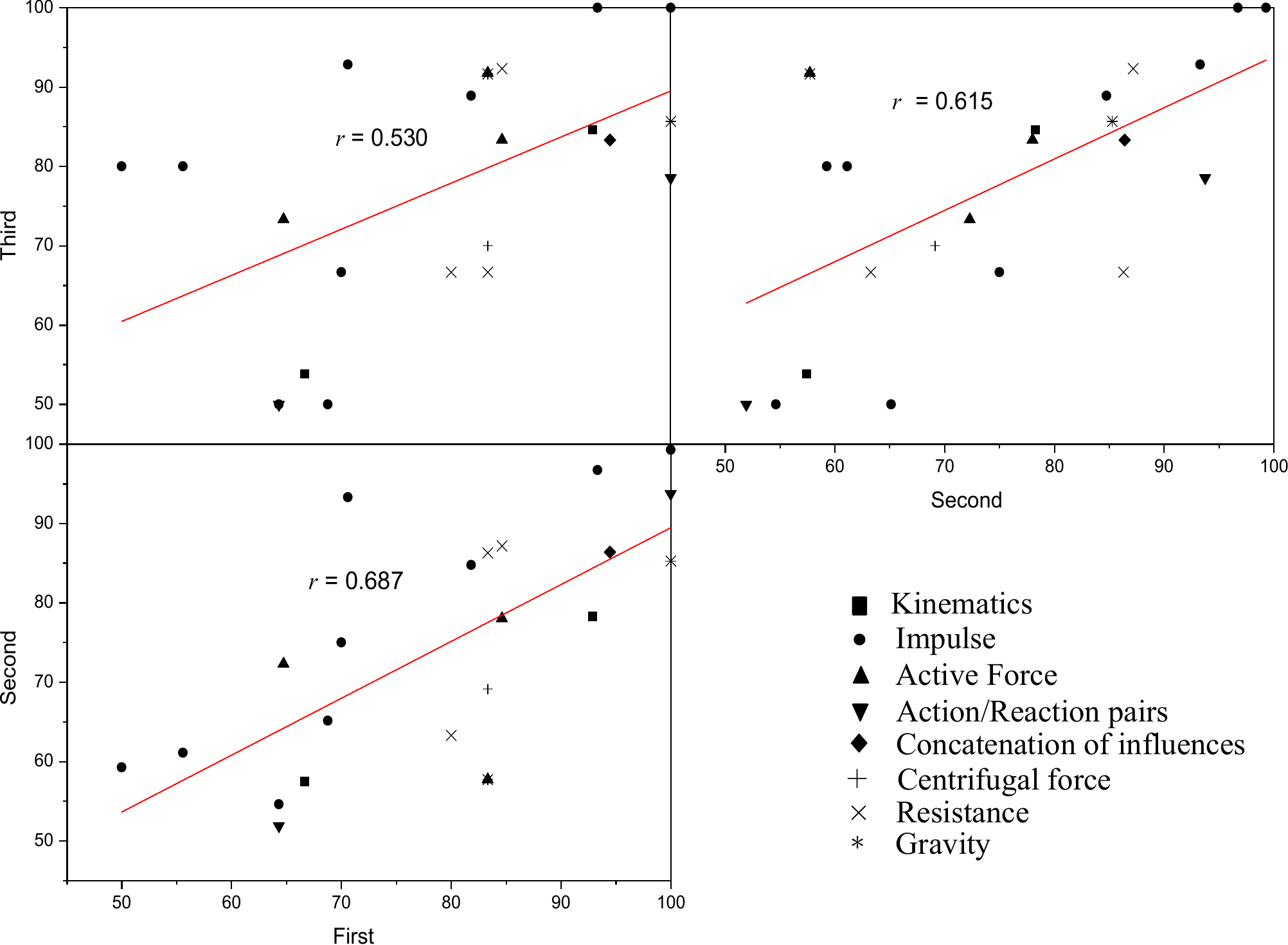}
    \caption{Correlation plot of the Phys. 1 course of the dominant incorrect answers between attempts.}
    \label{introcorr}
\end{figure}

Figure~\ref{introcorr} shows a plot between columns 3, 4, and 5 of table~\ref{desgloseI}. As we can see, there exists a positive correlation between attempts considered. The highest correlations are between the students in their first and second attempt, with a Pearson correlation factor $r=0.687$. We also see that the sections whit more highest percentages are, Impetus, specifically the misconceptions I1 associated with question 22 and I3 associated with question 5. Also, we found, in the Action/Reaction pairs section, AR1 associated with question 2, CI1 with question 18, and finally G3 with question 3.

\begin{figure}
    \centering
    \includegraphics[width=0.6\textwidth]{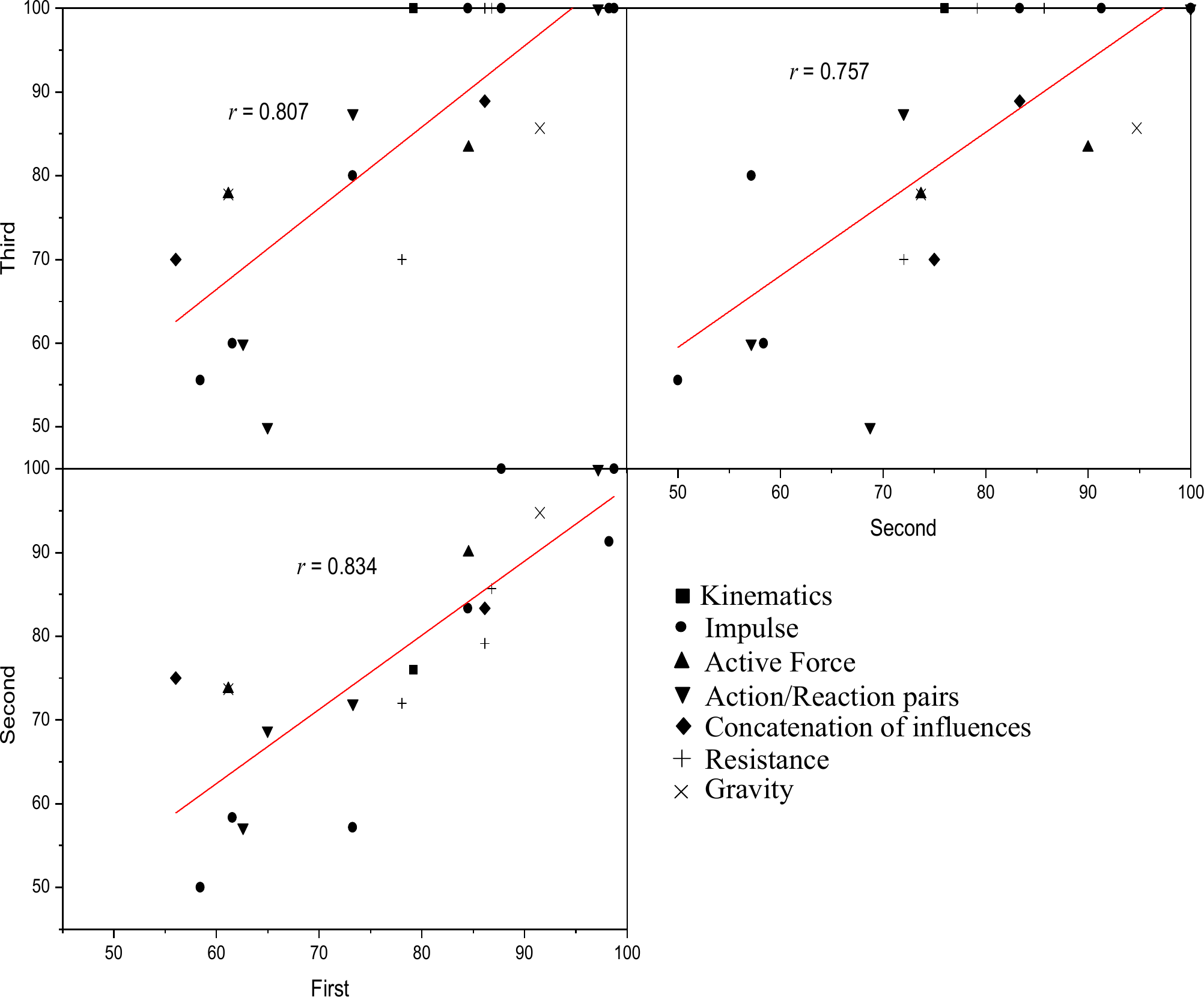}
    \caption{Correlation plot of the Phys. 2 course of the dominant incorrect answers between attempts.}
    \label{mecacorr}
\end{figure}

In figure~\ref{mecacorr} the correlations are all positives and higher than in Phys. 1 course. In this case, the highest Pearson correlation factor is $r=0.834$ and corresponds to the relation between the students in their first and second attempt. The misconceptions that have the highest percentages correspond to Impetus, Action/Reaction pairs, and Gravity. In the Impetus section, we found the questions 9, 22, and 16, associated with the misconceptions I1 and I3, in the Action/Reaction pairs section, we found the misconceptions AR1 associated with question 2, and G3 associated with question 3.

\begin{figure}
    \centering
    \includegraphics[width=0.6\textwidth]{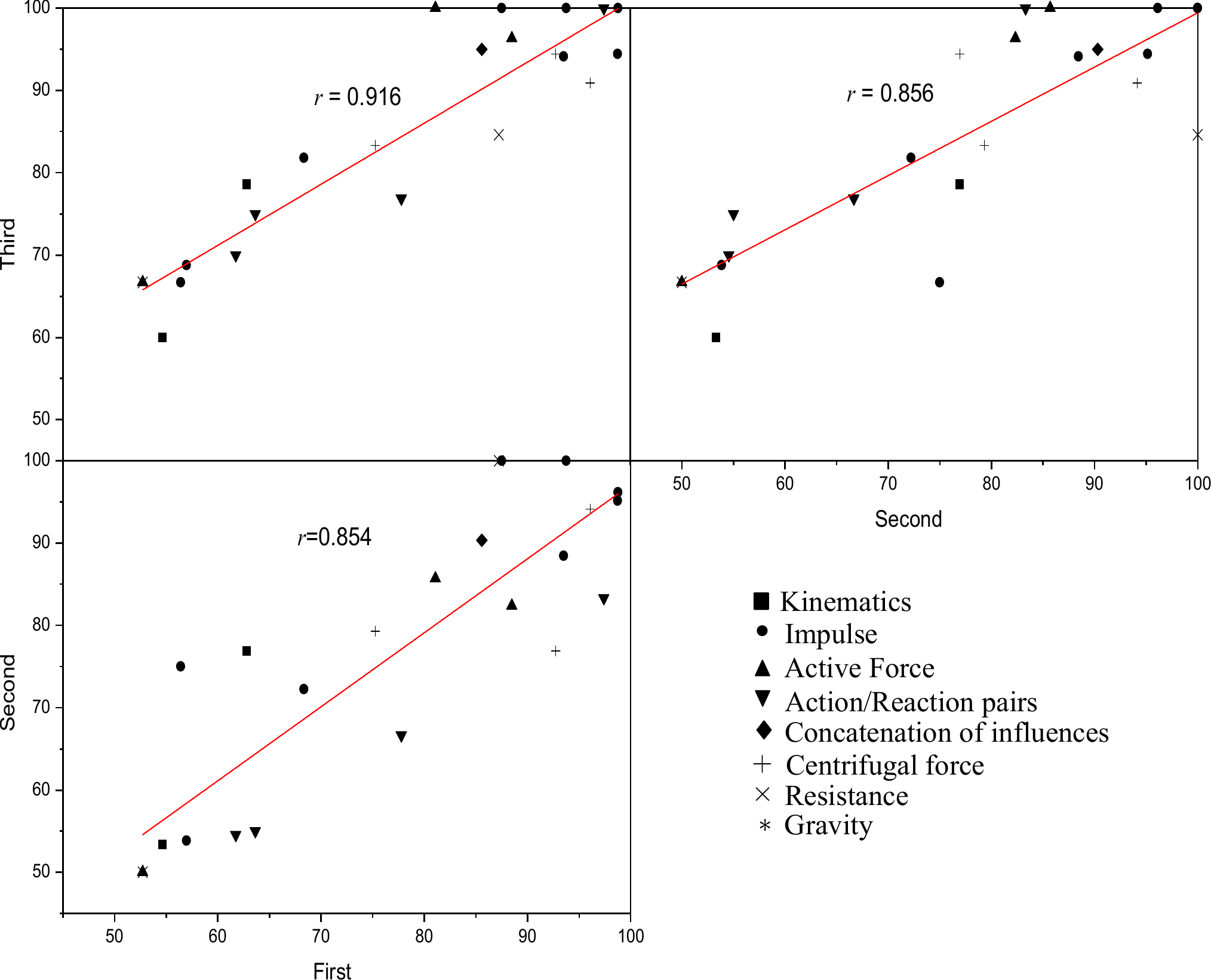}
    \caption{Correlation plot of the Phys. 3 course of the dominant incorrect answers between attempts.}
    \label{dinacorr}
\end{figure}

On the other hand, in figure~\ref{dinacorr}, as in the previous cases, the correlations are all positives. We found the highest values are between students in their first and third or more attempt with a Pearson correlation factor of $r=0.916$. The sections with the higher percentages are Impetus, Active Force, Action/Reaction pairs, Concatenation of Influences, and Other Influences on Motion. In the Impetus section, questions 9 and 22 are related to misconception I1, questions 5 and 16 with misconception I3, and questions 10 with misconception I5. In the Active Force section, we found the questions 12 y 15, associated with the misconception AF1. In the Action/Reaction pairs section, we found question 2 associated with the misconception AR1. Also, question 18, which corresponds to Concatenation Influences section, it's associated with the misconception CI1, and finally, we found questions 23 and 29 in the Other Influences on Motion section, associated with R1.

\begin{table*}
\centering
  \caption{Percentage of the dominant incorrect answer associated to their corresponding misconception code, classified by attempts, in Phys. 1 course.}
\begin{tabular}{lllll}
\toprule
 Question & Misconception code & First (\%) & Second (\%) & Third or more (\%) \\ \hline
20	&	K1	&	92.9	&	78.3	&	84.6	\\
21	&	K2	&	66.7	&	57.5	&	53.9	\\
9	&	I1	&	70.6	&	93.3	&	92.9	\\
22	&	I1	&	93.3	&	96.8	&	100.0	\\
26	&	I2	&	68.8	&	65.1	&	50.0	\\
5	&	I3	&	100.0	&	99.3	&	100.0	\\
16	&	I3	&	81.8	&	84.8	&	88.9	\\
27	&	I3	&	55.6	&	61.1	&	80.0	\\
8	&	I4	&	64.3	&	54.6	&	50.0	\\
4	&	I5	&	50.0	&	59.3	&	80.0	\\
10	&	I5	&	70.0	&	75.0	&	66.7	\\
12	&	AF1	&	64.7	&	72.3	&	73.3	\\
15	&	AF1	&	84.6	&	78.0	&	83.3	\\
17	&	AF5	&	83.3	&	57.7	&	91.8	\\
2	&	AR1	&	100.0	&	93.8	&	78.6	\\
13	&	AR2	&	64.3	&	51.9	&	50.0	\\
18	&	CI1	&	94.4	&	86.4	&	83.3	\\
4	&	CF	&	83.3	&	69.1	&	70.0	\\
23	&	R1	&	84.6	&	87.2	&	92.3	\\
29	&	R1	&	83.3	&	86.3	&	66.7	\\
28	&	R2	&	80.0	&	63.3	&	66.7	\\
3	&	G3	&	100.0	&	85.3	&	85.7	\\
17	&	G4	&	83.3	&	57.7	&	91.7	\\
 \bottomrule
\end{tabular}
\label{desgloseI}
\end{table*}

\begin{table*}
\center
  \caption{Percentage of the dominant incorrect answer associated to their corresponding misconception code, classified by attempts, in Phys. 2 course.}
\begin{tabular}{lllll}
\toprule
 Question & Misconception code & First (\%) & Second (\%) & Third or more (\%) \\ \hline
20	&	K1	&	79.2	&	76.0	&	100.0	\\
9	&	I1	&	98.3	&	91.3	&	100.0	\\
22	&	I1	&	98.7	&	100.0	&	100.0	\\
16	&	I3	&	87.8	&	100.0	&	100.0	\\
27	&	I3	&	73.3	&	57.1	&	80.0	\\
8	&	I4	&	58.4	&	50.0	&	55.6	\\
4	&	I5	&	61.5	&	58.3	&	60.0	\\
10	&	I5	&	84.5	&	83.3	&	100.0	\\
12	&	AF1	&	84.5	&	90.0	&	83.3	\\
17	&	AF5	&	61.2	&	73.7	&	77.8	\\
2	&	AR1	&	97.2	&	100.0	&	100.0	\\
11	&	AR1	&	73.3	&	72.0	&	87.5	\\
11	&	AR2	&	73.3	&	72.0	&	87.5	\\
13	&	AR2	&	62.6	&	57.1	&	60.0	\\
14	&	AR2	&	65.0	&	68.8	&	50.0	\\
18	&	CI1	&	86.1	&	83.3	&	88.9	\\
7	&	CI3	&	56.0	&	75.0	&	70.0	\\
23	&	R1	&	86.1	&	79.2	&	100.0	\\
29	&	R1	&	86.8	&	85.7	&	100.0	\\
28	&	R2	&	78.1	&	72.0	&	70.0	\\
3	&	G3	&	91.5	&	94.7	&	85.7	\\
17	&	G4	&	61.2	&	73.7	&	77.8	\\
 \bottomrule
\end{tabular}
\label{desgloseM}
\end{table*}

\begin{table*}
\center
  \caption{Percentage of the dominant incorrect answer associated to their corresponding misconception code, classified by attempts, in Phys. 3 course.}
\begin{tabular}{lllll}
\toprule
 Question & Misconception code & First (\%) & Second (\%) & Third or more (\%) \\ \hline
20	&	K1	&	62.8	&	76.9	&	78.6	\\
21	&	K2	&	54.7	&	53.3	&	60.0	\\
9	&	I1	&	93.5	&	88.5	&	94.1	\\
22	&	I1	&	98.7	&	95.2	&	94.4	\\
26	&	I2	&	57.0	&	53.9	&	68.8	\\
5	&	I3	&	98.8	&	96.2	&	100.0	\\
16	&	I3	&	93.8	&	100.0	&	100.0	\\
27	&	I3	&	68.3	&	72.2	&	81.8	\\
4	&	I5	&	56.4	&	75.0	&	66.7	\\
10	&	I5	&	87.5	&	100.0	&	100.0	\\
12	&	AF1	&	88.5	&	82.4	&	96.3	\\
15	&	AF1	&	81.1	&	85.7	&	100.0	\\
17	&	AF5	&	52.7	&	50.0	&	66.7	\\
2	&	AR1	&	97.4	&	83.3	&	100.0	\\
11	&	AR1	&	77.8	&	66.7	&	76.9	\\
11	&	AR2	&	77.8	&	66.7	&	76.9	\\
13	&	AR2	&	61.7	&	54.6	&	70.0	\\
14	&	AR2	&	63.6	&	55.0	&	75.0	\\
18	&	CI1	&	85.6	&	90.3	&	95.0	\\
23	&	R1	&	92.8	&	76.9	&	94.4	\\
29	&	R1	&	96.1	&	94.1	&	90.9	\\
28	&	R2	&	75.3	&	79.3	&	83.3	\\
3	&	G3	&	87.2	&	100.0	&	84.6	\\
17	&	G4	&	52.7	&	50.0	&	66.7	\\
 \bottomrule
\end{tabular}
\label{desgloseD}
\end{table*}

\section{Discussion}

In general terms, the FCI performance is low if we consider that 96.8\% of students who participated in this study have at least one attempt in the courses associated with Classical Mechanics. 
Considering that, the courses of this study are consecutive and related to each other, which means that a student in Phys. 3 course should have passed the Phys. 1 and Phys. 2 courses, so it has reviewed the topics of newtonian mechanics in these two courses, a fact that catches our attention, at first sight, is that there are just 4.8 points of difference between the mean score of Phys. 1 students on their first attempt and Phys. 3 students on their third attempt. 

According to Savinainen et al.\cite{Savinainen} an FCI score equal to 60\% (equivalent to 17.4 points in the 29 questions version) is the `` entry threshold '' to Newtonian physics, and below this value, the student's comprehension of newtonian physics is not enough, which is the case of the students of this study.

It is important to note that more than 50\% of questions are related to dominant incorrect answers, which are associated with their corresponding misconception defined by Hestenes et al.~\cite{FCI}. As we can see from figures (\ref{introcorr}), (\ref{mecacorr}) and (\ref{dinacorr}), all correlations are positive.

In the case of the Phys. 1 course, the highest correlation corresponds between students on their first and second attempt, while there is a weak correlation between students on their first and third attempt. On the other hand, in Phys. 2 course the correlations are greater than Phys. 1 course, and the highest correlation corresponds between students in their first and second attempt, while the weaker correlation is between the students in their second and third attempt. 
Finally, in the Phys. 3 course, the correlations are all high, and greater than Phys. 1 and Phys. 2 courses. There is a strong correlation between students on their first and third attempt. 

All defined misconceptions by Hestenes et al.~\cite {FCI} are present as dominant incorrect answers in this study. Also we note, from tables \ref{desgloseI}, \ref{desgloseM} and \ref{desgloseD} that the most dominant section is Impetus, whose misconceptions are I1, I3 and I5. For the case of I1, the percentage of choice between attempts is very high, above 70\% of students who answer incorrectly have the belief that there exists a motor force (impetus) that keep things moving, for example, a hit given to an object supply a force which is maintain acting throughout the movement. This is a pre Galilean conception that comes from an Aristotelian thought of the movement.

On the other hand, the misconception I3 shows that the students have the misconception that there must exist a motor force acting on to keep things moving, otherwise, there's no motion. For the case of misconception I5, we can infer that students believe that there exists a motor force that tends to move the objects in circular trajectories, namely, an object which moves in a circular trajectory tied to a rope, once released, will keep moving in a circular trajectory.

In the case of Phys. 2 and Phys. 3 courses, other repeating section is Action/Reaction pairs. Specifically, AR1 and AR2. The students tend to think that when two bodies interact with each other, which have more mass or are in movement, tend to dominate. This belief shows a conflict with the third Newton law.

\section{Conclusions}
Each course, independently, shows a similar mean score, which means that there is no big difference in overall student's performance in the FCI, according to the attempts. However, it's important to note that, if we compare the mean score between the three courses, this value has a slight increase. This can be explained by the fact that the students in Phys. 3 course,  specifically in their third attempt, has a better background in newtonian mechanics, provided in the two previous attempts in Phys. 1 and Phys. 2 courses, along with their curriculum.

However, if we focus only on the dominant incorrect questions, the scenario changes. This means that, if we consider the choice of misconceptions, there is no big difference between a student on their third attempt and a student on their first attempt in each course.

Based on the correlations between attempts, in each curse, and considering that all courses of the sample have never undergone major modifications, at least since that a student, currently in Phys. 3, was in Phys. 1 course, the results suggest that the number of attempts isn't an independent variable. It may be affected by factors such as previous courses that contribute to their background (like calculus courses) and their abilities (e.g. problem-solving skills, scientific reasoning, etc.) or in their approaches and habits.

On the other hand, students tend to present confusion between acceleration and force in the context of an object falling by the action of gravitational force. Most of the students tend to think that the acceleration of gravity increase significantly close to the surface of the Earth. This tendency observed is decreasing if we compare students who are in the Phys. 1 and Phys. 3 courses. Predominantly, the students show confusion with the concept of Impetus and the third Newton law.

Finally, it is precise to emphasize that the results of this study focus only on the choice of misconceptions, considering their attempts, in each course independently. There is suggested that for further works, it consider other variables mentioned before.

\end{document}